\documentclass[twocolumn,aps,prb,groupedaddress,amssymb,showpacs]{revtex4}
\usepackage{graphicx}
\usepackage{dcolumn}
\usepackage{amsmath}
\begin{document}

\title{Far field imaging by a planar lens: diffraction versus superresolution }


\author{Nicholas A. Kuhta}
\author{Viktor A. Podolskiy}
\email{vpodolsk@physics.oregonstate.edu}
\affiliation{Department of Physics, 301 Weniger Hall, Oregon State
University, Corvallis OR 97331 USA}

\author{Alexei L. Efros}
\email{efros@physics.utah.edu} \affiliation{Department of Physics,
University of Utah, Salt Lake City UT, 84112 USA}


\begin{abstract}
We resolve the long standing controversy regarding the imaging by a
planar lens made of left-handed media and demonstrate theoretically
that its far field image has a fundamentally different origin
depending on the relationship between losses {inside} the
lens and the wavelength of the light $\lambda$. At small enough
$\lambda$ the image is always governed by diffraction theory,
and the resolution is independent of the absorption if both
Im$\epsilon \ll 1$ and Im$\mu \ll 1$. For any finite $\lambda$,
however, a critical absorption exists below which
the superresolution regime takes place, though this absorption is
extremely low and can hardly be achieved. We demonstrate that the
transition between diffraction limited and superresolution regimes
is governed by {the} universal parameter combining
absorption, wavelength, and lens thickness. Finally, we show that this
parameter is  related to the resonant excitation of the surface
plasma waves.

\end{abstract}
 \maketitle

The left-handed medium (LHM), introduced  by Veselago\cite{ves} in
1967 as a medium with simultaneously negative and real $\mu$ and
$\epsilon$ provides a  negative refraction at its interface with a
regular medium (RM). This effect allows creation of a unique imaging
device, sometimes called the ``Veselago lens'', formed by a planar
slab of LHM, with its refractive index and impedance ideally matched
to the surrounding RM (Fig.1). Interest in the planar lens
significantly increased after the work by Pendry\cite{pendryo}, who
suggested that the planar lens can in principle focus all Fourier
components of a 2D image, introducing a term ``the perfect lens''.
However, taken literally, this statement contradicts Electrodynamics: since there is no source at the focal point the field of the source cannot be exactly reproduced in the vicinity of the focal point. As it was
pointed out in Refs.[\onlinecite{hal, efros1,garsia}], in the
absorption-less limit (${\rm Im}\epsilon={\rm Im}\mu=0$), the
solution proposed in\cite{pendryo} exponentially diverges inside a
3D domain between two foci (see Fig.1), and therefore it cannot be a
solution of Maxwell's equations. Pokrovsky and Efros\cite{efros1,
efros2} proposed a diffraction theory that should be valid at large
$k_0=\omega/c$ and that does not contain any superresolution.
Haldane\cite{hal} explained that the superresolution should be
connected to the resonance of the plasma waves on both sides of the
slab, but he added that the theory should be regularized. The
regularization due to a very small absorption inside {the} LHM, and
its implication for resolution {were} analyzed by {many}
authors\cite{smith,pn,milton2,merlin,shvetsResolut,webb}. Near field
behavior has been studied since 1994\cite{nicorovici,milton1}, both theoretically and experimentally\cite{zhangSuperlens,blaikie,shvetsExperiment,smolyaninov}. The goal of this work is to develop a quantitative
criterion of applicability of diffraction theories for planar lens
systems.

\begin{figure}[t]
\includegraphics[width=6.cm]{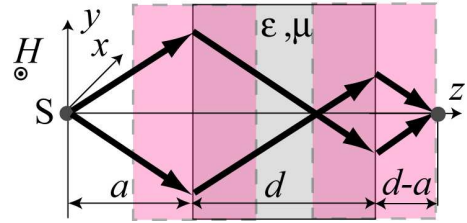}
\caption{(Color online)Planar LHM-based lens with two real foci outside and
inside the slab. S is the source, $a$ is the distance from the
source to the slab, $d$ is the width of the slab. {Dark (Red)
regions with dashed boundaries are the } resonant regions. The fields diverge in these regions when the absorption tends to zero.
\label{Fig1} }
\end{figure}

It has been shown that the far-field imaging of the planar lens has
the following properties:

 (a) The stationary solution of the Maxwell equations does
not exist when Im$\epsilon$=Im$\mu$=0. When absorption is small, but
non-zero, there are two ``resonant" regions shown in Fig. 1; the
fields inside resonant regions diverge as absorption tends to zero.
Both foci are at the boundaries of the resonant regions. Therefore
the focus outside the lens is quasi virtual: an observer to the
right of the focal point can see subwavelength focus while an
observer to the left of the focus can see only large and highly
oscillating fields -- an indication of resonantly excited
surface plasma waves propagating at the back interface of the lens. The
existence of such a quasi-focus does not contradict general
theorems.

 (b) Milton {\it et al.}\cite{milton2} showed that total
absorption of electric energy inside the slab $\propto({\rm
Im}\;\epsilon)^{(2a/d-1)}$, where $a$ is the distance from the
source to the lens and $d$ is the width of the lens. If $a<d/2$, the
resonant regions overlap. In this case {\em the total absorption
inside the lens tends to infinity as {\rm Im}$\epsilon$ tends to
zero.}  Moreover, the electrostatic lens cloaks two dimensional
dipoles rather than imaging them ``perfectly" (See also Ref.[\onlinecite{enghetaCloak}]).

On the other hand, 2D and 3D diffraction theories demonstrate a real
focus with different widths in the lateral and in the longitudinal
directions. According to these theories, fields near the focus are
the universal function of ${\bf r}/\lambda$, where ${\bf r}$ is the
radius-vector with an origin at the focal
point\cite{efros1,efros2,lhe}. Although diffraction theory uses
spherical waves rather than plane waves, results of both 2D and 3D
diffraction theories coincide with solutions in the plane wave
representation with evanescent waves omitted. The diffraction theory
is known to be the first approximation at small deviations from
geometrical optics\cite{ lan, jac}. However in the case of the
planar lens we have a paradoxical situation. If Im$\epsilon\ll 1$
and Im$\mu\ll 1$, the width of the focus, as obtained by the
diffraction theory, is independent of absorption, while the exact
solution has a singularity at zero absorption even at small
wavelengths. Therefore, the criterion of applicability of the
diffraction theory cannot be related to a trivial ratio of
wavelength to width of the slab, and must also depend on absorption.

Podolskiy and Narimanov\cite{pn} studied the width of the focus of
the 2D planar lens. They have found that it is close to the
diffraction limit almost at all cases where the wavelength is
smaller than the width $d$ of the LHM slab. They also made an
important observation that the boundaries  of the diffraction
regions depend on the absorption.

First we describe our results.   We  consider TM polarization with a
2D Green's function as a source. The magnetic field  with an amplitude
$h$ near the source at the origin has a form $G_R=(i/4)h H_0^{(1)}(\rho k_0)$,
 with $H$ being the Hankel function $H^{(1)}_0=J_0+iN_0$, and $\rho=\sqrt{z^2+y^2}$.
 This function can be represented in the  form $G_R=H_p+H_{ev}$, where

\begin{equation}
H_p=\frac{ih}{\pi}\int_{-k_0}^{k_0} {\exp{i\left(k
y+z\sqrt{k_0^2-k^2}-\omega t\right)}\over \sqrt{k_0^2-k^2}}dk
\label{p}
\end{equation}
contains only propagating waves while
\begin{equation}
H_{ev}=\frac{h}{\pi}\int_{|k|>k_0} \frac{\exp\left(ik
y-z\sqrt{k^2-k_0^2}-i\omega t\right)}{\sqrt{k^2-k_0^2}}dk \label{ev}
\end{equation}
contains only evanescent waves (EW).

The 2-dimensional diffraction theory\cite{lhe,vd} shows that the
field near the focus contains only propagating modes. If the slab
is at $a<z<d+a$ and  Im$\epsilon \ll 1$ and Im$\mu \ll 1$, the field has
a form
\begin{equation}
 H_p^f(y,z^\prime)=\frac{ih}{\pi}\int_{-k_0}^{k_0} {\exp{i\left(k y+z^{\prime}\sqrt{k_0^2-k^2}-\omega t\right)}\over
\sqrt{k_0^2-k^2}}dk,
\end{equation}
where $z^\prime=z-(2d-a)$ is a distance from the focus in
z-direction. At $z^\prime=0$ one gets $ H_p^f(y,0)=ihJ_0(2\pi
y/\lambda)$ Note that the halfwidth of the first maximum of the
square of the  Bessel function is $\Delta y\approx  0.3\lambda$.
This small value of the width in the lateral direction has been
misinterpreted as
superlensing\cite{fanMultilayers,podolskiyComment}.

As a criterion of transition from the supersolution regime to the
diffraction regime we choose the ratio
$P=H_{ev}/H_p^f(0,0)=H_{ev}/ih$, where both magnetic fields  of EW's
and magnetic fields  given by the diffraction theory are taken at the
focal point $z=2d-a$. Then if $P$ is small, we have the diffraction
regime, while at large $P$ we have the  regime of supersolution.

One should keep in mind that the creation of the LHM is a difficult
and controversial problem. Usually the LHMs are metamaterials with
periodic or quasi-periodic structure\cite{shalaev} and their
magnetic permeability $\mu$ may exhibit strong spatial dispersion
[$\mu=\mu(\omega,\vec{k})$] \cite{efrospc,lhe}.  All these problems
are outside the scope of this paper.  Here we consider the basic
 model -- a homogeneous and isotropic  ``hypothetical" slab that has
\begin{equation}\label{eps}
\epsilon=\mu=-1+i\delta,
\end{equation}
 where $\delta \ll 1$;
$\epsilon=\mu=1$ outside the slab. Our goal is to show what one can expect in the best case scenario.

\begin{figure}[t]
\includegraphics[width=8cm]{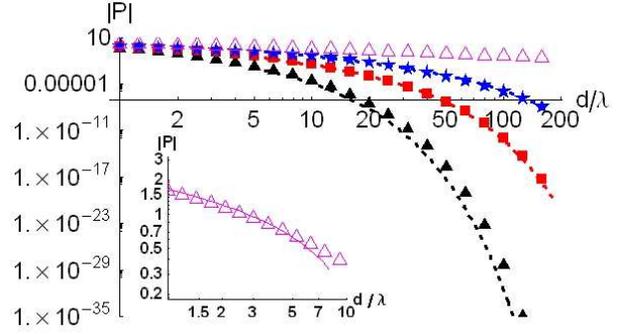}
\caption{ (color online) Dependence of the parameter $|P|$ on
normalized lens thickness for $\delta=10^{-2}$ (solid triangles),
$\delta=10^{-3}$ (boxes),  $\delta=10^{-4}$ (stars), and $\delta=10^{-18}$ (empty triangles); symbols correspond to results of numerical integrations; dashed and solid lines correspond to approximate Eq.(\ref{gS}) and Eq.(\ref{lS}) respectively; note that quasistatic Eq.(\ref{lS}) adequately describes the far-field behavior of the system in the limit of vanishingly small absorption
\label{FigPS} }
\end{figure}

Our main results are:

(i) The parameter responsible for the far-field ($k_0 d\gg 1$) transition between diffraction limited and
 superresolution regimes is not described by the wavelength alone, but also depends on absorption:
\begin{equation}\label{S}
  S=k_0 d \sqrt{{\rm Im}(\epsilon+\mu)/2}=k_0 d \sqrt{\delta}.
\end{equation}

(ii) At $S \gg 1$ the system is always described by diffraction theory, and
\begin{equation}\label{gS}
P(S)=\frac{1}{k_0 d}\frac{4 (i\sin{S}-\cos{S})\exp({-S})}{ S(i-1)}\ll 1.
\end{equation}

(iii) At $S < 1$ (and $k_0 d>1$) the system may exhibit superresolution, with
\begin{equation}\label{mS}
P(S)=\frac{1}{i\pi}\sinh^{-1}(y_f/(2k_0 d)),
\end{equation}
 where $y_f$ is  the root of
equation
\begin{equation}\label{yf}
\frac{4 S^4 \exp{y_f}}{y_f^4}=1.
\end{equation}

(iv) Finally, when $-\ln(\delta/4)\gg k_0 d$ ($S \ll 1$) the system exhibits superresolution with
 quasistatic-like behavior
\begin{equation}\label{lS}
P(S)\simeq\frac{2}{i\pi}\ln\left[-\frac{2 \ln(\delta/4)}{k_0 d}\right] \propto \ln\ln(1/S).
\end{equation}
$P(S)\rightarrow \infty$ as $S\rightarrow0$. However, divergence of $P(S)$ is extremely slow, and in practice $|P(S)|\gtrsim 1$ is unachievable in realistic far-field structures.

Note that the limit $\lambda\rightarrow 0$ ($k_0\rightarrow\infty$)
at a fixed $\delta$ and the limit  $\delta\rightarrow 0$ at fixed
$k_0 (\lambda)$ do not commute. Indeed, at small absorption the
diffraction regime exists at large enough $k$ and that at large $k$
the superresolution regime exists at small enough absorption; as
described above, the solution at $\delta=S=0$ does not exist. Below
we provide the derivation of
 Eqs.(\ref{S},\ref{gS},\ref{mS},\ref{lS}), and demonstrate the connection between
 the superresolution and interaction of plasmonic surface waves, which disappears at $S>1$.

To derive Eqs. (\ref{p},\ref{S},\ref{lS}) we use transmission
coefficients for EW's as calculated  by Podolskiy and
Narimanov\cite{pn} and take into account that $H_p^f(0,0)=ih$. Then
\begin{widetext}
\begin{eqnarray}
  P
  &=&\frac{2}{i\pi}\int_{k_0}^\infty\frac{\exp({-d\sqrt{k^2-k_0^2}})dk}{\sqrt{k^2-k_0^2}[D
  \sinh{(\sqrt{k^2-k_0^2\epsilon \mu}d})+\cosh{(\sqrt{k^2-k_0^2\epsilon \mu}d)}]}
\label{intExact}
\\
  D&=& -\frac{k_0^2\epsilon(\epsilon+\mu)-k^2(1+\epsilon
  ^2)}{2\epsilon\sqrt{k^2-k_0^2}\sqrt{k^2-k_0^2\epsilon \mu}}
\nonumber.
\end{eqnarray}

At $k_0 d \gg 1$ the values of $k$, that are important in this
integral are very close to $k_0$. Namely,
\begin{equation}\label{est}
k^2-k_0^2\sim 1/d^2.
\end{equation}
Then $(k-k_0)/k_0\sim 1/d^2k_0^2\ll 1$, $k^2-k_0^2\sim 2k_0(k-k_0)$, and $\epsilon^2=\epsilon\mu\approx 1-2i\delta$ [see Eq.(\ref{eps})]. Thus,
\begin{equation}\label{int}
P=-\frac{2}{\pi k_0 d}\int_0^\infty \frac{\exp({-y})
dy}{\sqrt{1+iS^2/2y^2}\sinh{\sqrt{y^2+iS^2/2}}+\cosh{\sqrt{y^2+iS^2/2}}}.
\end{equation}
\end{widetext}
As it can be explicitly verified, in the regime $S\gg 1$, the denominator of this integral can be further simplified:
$\sqrt{1+iS^2/2y^2}\sinh{\sqrt{y^2+iS^2/2}}+\cosh{\sqrt{y^2+iS^2/2}}\simeq \sqrt{i S^2/2y^2}\sinh{\sqrt{iS^2/2}}$, leading to Eq.(\ref{gS}).

If $S=0$ the integral in Eq.(\ref{int}) diverges which means that
there is no solution without absorption. To consider the case of
small $S$ it is more convenient to start with the transmission given
by Eq.(2) of Ref.\cite{pn} that is written for the case of small
absorption. Introducing a new variable of integration
$y=2\sqrt{k^2-k_0^2}d$ one gets
\begin{equation}\label{C}
P=\frac{1}{\pi k_0d}\int_0^\infty\frac{dy}{\sqrt{(y/(2 k_0
d)^2+1}(1+\phi^2\exp(y))},
\end{equation}
where
\begin{equation}\label{fi}
\phi(y)=\frac{1}{2}\left({\rm Im} \epsilon+\frac{4\delta (k_0 d)^2}{y^2}\right)
\end{equation}
In the limit of extremely small absorption, this integral is dominated by the first term in Eq. (\ref{fi}),
 yielding a quasi-electrostatic-like regime [Eq.(\ref{lS})], predicted by Nicorovici {\it et
al.}\cite{nicorovici}, where the results are independent of $\mu$
Note that {\it this quasi-electrostatic
 regime may exist even at $k_0d\gg 1$ if absorption is small enough.}

For somewhat larger absorptions, the far field regime is described by the second term in Eq.(\ref{fi}), leading to
\begin{equation}\label{C1}
P=\frac{1}{\pi k_0d}\int_0^\infty\frac{dy}{\sqrt{(y/(2k_0
d)^2+1}(1+4S^2 e^{y}/y^4)}.
\end{equation}
At $S\ll 1$ one gets an adequate approximation for the integral
Eq(\ref{C1}) assuming that $1/[1+4S^2\exp(y)/y^4]=\vartheta(y_f-y)$,
where $\vartheta(x)$ is the Heaviside Step function and $y_f$ is given by Eq.(\ref{yf}). This leads to Eq.(\ref{mS}).

Fig.\ref{FigPS} shows the comparison of our analytical results for $P(S)$ for $S\ll 1$ and $S\gg 1$ to the results of numerical integration of Eq.(\ref{intExact}).

We now explain the physical meaning of parameter $S$. The condition
$k_0 d \gg 1$ is not sufficient for the applicability of diffraction
theory in the planar  lens because of the resonance interaction of
the surface  plasmon waves\cite{hal}. These waves exist in TM
polarization under the condition  \cite{lan,rup,hal} that the ratio $k_z/\epsilon$ changes sign at the interface of the vacuum and the LHM. At
$\delta\ll 1$ it reads
\begin{equation}\label{sw2}
\sqrt{k^2-k_0^2}=\sqrt{k^2-k_0^2+k_0^2 2i\delta}.
\end{equation}
The mismatch responsible for the breakdown of resonant excitation of
surface modes can be related to the term  $k_0^2 2i\delta$. At $S\gg
1$ one can use Eq.(\ref{est}) to find that the absolute value of the
ratio of the mismatch term to $k^2-k_0^2$ is $S^2\gg 1$. Thus, at
$S\gg 1$ there are no traces of the resonance and regular
diffraction theory should be  applicable. At $S\ll 1$ it follows
from Eq. (\ref{C}) that $k^2-k_0^2\sim y_f^2/d^2$. Ignoring in this
estimate the logarithmic factor in $y_f$ and assuming $y_f\sim 1$ we
resolve that the relative mismatch in Eq. (\ref{sw2}) is also of the
order of $S^2$. Therefore, the violation due to the absorption is
small at $S\ll 1$ and the main features of the resonance should be
preserved.

\begin{figure}
\includegraphics[width=7cm]{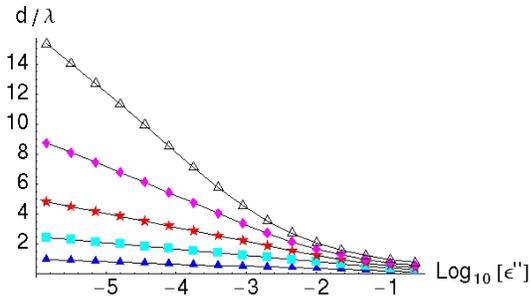}
\caption{ (color online) The normalized lens thickness as a function
of absorption for constant values of parameter $|P|$; filled
triangles, squares, starts, diamonds, and empty triangles correspond
to $|P|=1$, $|P|=1/2, |P|=1/4, |P|=1/8$, and $|P|=1/16$
respectively. In the part of the plane above the curve with a given
$|P|$, the relative  contribution of evanescent waves is less than
$|P|$.
 \label{Fig2} }
\end{figure}

Finally, we use the developed formalism to analyze the range of
parameters where one could expect superresolution (and quasi virtual
focus with subwavelength thickness). These results are summarized in
Fig.\ref{Fig2} which shows the dependence of normalized lens thickness
$d/\lambda$ as a function of absorption $\delta$ for a set of fixed
values of the parameter $|P|$, calculated using direct numerical
calculation of the integral in Eq.(\ref{intExact}). One can see that the
 contribution of the evanescent waves at the focal point is
 practically negligible at $d/\lambda >2$ at any reasonable value
 of absorption.

 In conclusion, we have developed an approach to
calculate a quantitative measure of superlensing in a planar
hypothetical  LHM-based lens, and used the developed formalism to
separate the regions of the superresolution and diffraction in the
far field regime. We demonstrated that the limits of absorption
$\delta \rightarrow 0$ and wavelength $\lambda \rightarrow 0$ do not
commute; the former limit yields superlensing, while the latter
leads to diffraction limited behavior, which typically dominates the
far-field image  of realistic planar lenses. We demonstrated that if
$\lambda$ and $\delta$ are both finite, the behavior of the planar
lens is described by a universal parameter $S$ which depends on {\it
both geometrical sizes and absorption}, and found analytically
asymptotical behaviors for $S \rightarrow 0$ and $S\rightarrow
\infty$. A connection between the value of $S$ and the existence of
resonant excitation of plasmonic waves has also been demonstrated.
Understanding the onset of the diffraction limit presented in our
work is important for the further development and design of
 imaging systems with negative refraction.

\bibliography{dif}

\end{document}